# Ultrafast ferromagnetic fluctuations preceding magnetoelastic first-order transitions

Zhao Zhang[1,2,#], Yanna Chen[3,4,#,*], Dehong Yu[5], Richard Mole[5], Chenyang Yu[1,2], Zhe Zhang[1,2], Houbo Zhou[6,7], Xuexi Yan[1,2], Xinguo Zhao[1], Weijun Ren[1], Chunlin Chen[1,2], Shigeki Owada[8,9], Kensuke Tono[8,9], Michihiro Sugahara[9], Yue Cao[10], Osami Sakata[4,8], Michael J. Bedzyk[3], Bing Li[1,2,*], Fengxia Hu[6,7,*], Baogen Shen[6,7], Zhidong Zhang[1,2]

[1] Shenyang National Laboratory for Materials Science, Institute of Metal Research, Chinese Academy of Sciences, 72 Wenhua Road, Shenyang, Liaoning 110016, China.
[2] School of Materials Science and Engineering, University of Science and Technology of China, 72 Wenhua Road, Shenyang, Liaoning 110016, China.
[3] Department of Materials Science and Engineering, Northwestern University, Evanston, Illinois 60208, USA.
[4] Synchrotron X-ray Group, Research Center for Advanced Measurement and Characterization, National Institute for Materials Science, 1-1-1 Kouto, Sayo-cho, Hyogo 679-5148, Japan.
[5] Australian Nuclear Science and Technology Organisation, Locked Bag 2001, Kirrawee DC NSW 2232, Australia.
[6] Beijing National Laboratory for Condensed Matter Physics and State Key Laboratory of Magnetism, Institute of Physics, Chinese Academy of Sciences, Beijing 100190, China.
[7] School of Physical Sciences, University of Chinese Academy of Sciences, Beijing 100049, China.
[8] Japan Synchrotron Radiation Research Institute, 1-1-1 Kouto, Sayo-cho, Sayo-gun, Hyogo 679-5198, Japan.
[9] RIKEN SPring-8 Center, 1-1-1 Kouto, Sayo-cho, Sayo-gun, Hyogo 679-5198, Japan.
[10] Materials Science Division, Argonne National Laboratory, Lemont, Illinois 60439, USA.
* Corresponding authors: bingli@imr.ac.cn, fxhu@iphy.ac.cn or yanna.chen@northwestern.edu.
# These authors equally contributed to the research.

**First-order magnetic transitions are of both fundamental and technological interest given that a number of emergent phases and functionalities are thereby created[1-5]. Of particular interest are giant magnetocaloric effects, which are attributed to first-order magnetic transitions and have attracted broad attention for solid-state refrigeration applications[6]. While the conventional wisdom is that atomic lattices play an important role in first-order magnetic transitions[7,8], a coherent microscopic description of the lattice and spin degrees of freedom is still lacking. Here, we study the magnetic phase transition dynamics on the intermetallic $LaFe_{13-x}Si_x$, which is one of the most classical giant magnetocaloric systems, in both frequency and time domains utilizing neutron scattering and ultrafast X-ray diffraction. We have observed a strong magnetic diffuse scattering in the paramagnetic state preceding the first-order magnetic transition, corresponding to picosecond ferromagnetic fluctuations. Upon photon-excitation, the ferromagnetic state is completely suppressed in 0.9 ps and recovered in 20 ps. The ultrafast dynamics suggest that the magnetic degree of freedom dominates this magnetoelastic transition and ferromagnetic fluctuations might be universally relevant for this kind of compounds.**

First-order magnetic transitions (FOMTs) take place with abrupt changes of several physical quantities at transition points with respect to temperature, magnetic field, electric field, pressure, stress, etc[1-5]. Such features offer a great opportunity for applications in a variety of functional sensors with high intrinsic signal-to-noise ratios. More importantly, the multiple degrees of freedom are often strongly coupled at FOMTs so that some unprecedented functionalities become possible based on crossing control. FOMTs can be categorized into magneto-structural and magnetoelastic transitions[9]. The former involves two magnetic phases accommodating different crystallographic symmetries and are interpreted based on soft-mode phonons and their coupling with magnetic moments[10]. By contrast, there is a discontinuity of crystal unit cell volumes found in the latter with the crystallographic symmetry unchanged. The frequently adopted mechanism for this case is the itinerant electron metamagnetism (IEM)[11]. An IEM transition, i.e., a FOMT from Pauli paramagnetic (PM) to ferromagnetic (FM) states in an itinerant magnetic system induced by external magnetic fields, is formulated based on the Landau-Ginzburg free energy expansion taking into account the renormalization effect associated with spin fluctuations. A sharp peak in electronic density of states below the Fermi level tends to render a negative fourth-order coefficient of the free energy expansion and an IEM is expected as demonstrated in many rare-earth and transition-metal compounds[11].

Among these compounds, $LaFe_{13-x}Si_x$ has attracted broad attention for its superior magnetocaloric performances[6,12]. With $x$ in a very narrow doping range around 1.5, the system undergoes a sharp FM-PM FOMT at Curie temperature ($T_C$) of about 190 K, where there is a strong magnetoelastic coupling as reflected in a dramatic negative thermal expansion ($\Delta V/V \sim 1.5\%$, where $V$ is the unit cell volume). Above $T_C$, a moderate magnetic field can induce a FOMT, giving rise to giant magnetic entropy changes of about -25 J kg$^{-1}$ K$^{-1}$. $T_C$ of this system can be elevated up to 350 K by substituting Fe with Co atoms or by adding interstitial hydrogen atoms, which hardly reduces the magnetic-entropy changes[13,14]. For instance, $La(Fe_{0.94}Co_{0.06})_{11.9}Si_{1.1}$ has maximum magnetic-entropy changes of -20 J kg$^{-1}$ K$^{-1}$ at $T_C \sim 280$ K[14]. Moreover, the thermal and magnetic hysteresis at the FOMT is small compared with other giant magnetocaloric materials[6]. In terms of real applications, $LaFe_{13-x}Si_x$ is also cost-effective given that it consists of earth-abundant elements.

Despite the outstanding magnetocaloric performances mentioned above that make this system promising for solid-state refrigeration applications, the exact mechanism of its FOMT remains unclear. A fundamental understanding of FOMT may foster the future designing of advanced magnetocaloric materials. IEM has been frequently used to account for the transition, but spin fluctuations themselves as the core of this mechanism have merely been investigated yet[15-17]. In recent neutron diffraction measurements, spin fluctuations arising from PM correlations were discovered above $T_C$[18]. In addition, nuclear resonant inelastic X-ray scattering shows partial phonon density of state (PDOS) of Fe sublattices is renormalized across $T_C$ in LaFe$_{11.5}$Si$_{1.5}$ (ref. 8). Thus spin-phonon coupling is believed to play a critical role in this system[8]. To thoroughly clarify the FOMT, it is necessary to conduct research considering lattice and magnetic degrees of freedom in both static and dynamic respects. Here, we present an investigation comparing LaFe$_{11.6}$Si$_{1.4}$ and LaFe$_{11.2}$Si$_{1.8}$, which exhibit a FOMT and a second-order magnetic transition (SOMT), respectively. Combining quasi-elastic neutron scattering (QENS) and ultrafast X-ray powder diffraction (XPD) based on X-ray free electron laser (XFEL), we are able to establish the differences between the two materials, and pinpoint the ultrafast FM fluctuations as the decisive factor in the FOMT and hence in the giant magnetocaloric effect.

LaFe$_{13-x}$Si$_x$ intermetallics crystalize in the cubic NaZn$_{13}$-typed structure with space group $Fm\bar{3}c$ [6]. As shown in **Fig. 1a**, the crystal structure appears as a cage configuration consisting of Fe atoms in which La atoms are trapped. There are two inequivalent crystallographic sites for Fe, i.e., Fe(I) on the eightfold (8$b$) and Fe(II) on the 96-fold (96$i$) Wyckoff positions. The Fe(II) site is shared with Si. LaFe$_{11.6}$Si$_{1.4}$ has a FOMT at 184 K, while a SOMT occurs in LaFe$_{11.2}$Si$_{1.8}$ at 208 K (see **Supplementary Fig. 1** for their temperature dependencies of magnetization). Given that the crystal structure is the same, the chosen compositions are an ideal playground for understanding the lattice and spin dynamics.

Shown in **Fig. 1b** and **c** are PDOS for these two compositions in the vicinity of transitions up to 50 meV. The profiles of PDOS of the two compositions both feature a broad peak located at about 24 meV with a few small local maxima superimposed, in agreement with the previous report[8]. The profiles are essentially dominated by the partial PDOS from the Fe atoms as determined in nuclear resonant inelastic X-ray scattering measurements[8]. The overall line shapes don't change significantly across $T_C$, except for the shift to lower energies upon becoming PM. To quantitatively characterize the phonon softening at the transitions, we track the peak at 24 meV as a function of temperature. As plotted in **Fig. 1d**, both compositions display remarkable softening. The decreases of energy at heating from 100 to 300 K are similar for both, but the change of LaFe$_{11.6}$Si$_{1.4}$ is slightly sharper, consistent with its nature of FOMT. Obviously, the magnetic softening of phonons across $T_C$ are not unique to FOMT, but also for SOMT. In this sense, phonon softening is impossible to solely account for differences between these two compositions. As such, in addition to phonons, we also examine the magnons. The FM states of both compositions show almost identical spin excitations below $T_C$. At 100 K, for example, strong spin excitations develop from the first Bragg peak, (200), and persist up to about 2 meV (**Supplementary Fig. 2**).

Hereafter, we turn our attention to the PM states where spin fluctuations are expected. In **Fig. 2a**, we plot the dynamic structure factor $S(Q,E)$ as a function of momentum transfer ($Q$) and energy transfer ($E$) at selected temperatures near $T_C$. In LaFe$_{11.6}$Si$_{1.4}$, at 210 K just above $T_C$, intense diffuse scattering is observed in the low-$Q$ region. The intensity goes weaker as temperature rises from 210 to 300 K, but pronounced diffuse scattering is still detected at 300 K. In LaFe$_{11.2}$Si$_{1.8}$, similar diffuse scattering is found (see **Supplementary Fig. 3**). In search for possible differences between them, we exclude the elastic intensity by integrating the spectra in the intervals of $-2.55 \leq E \leq -0.15$ meV and $0.15 \leq E \leq 2.55$ meV at about $1.2T_C$ (210 K for LaFe$_{11.6}$Si$_{1.4}$ while 250 K for LaFe$_{11.2}$Si$_{1.8}$), as plotted in **Fig. 2b**, given that the experimental energy resolution is about 0.135 meV. The obtained intensity of both compositions rapidly grows as $Q$ decreases when $Q < 0.8$ Å$^{-1}$, typical of PM scattering[19]. Strikingly, an extra intensity is found between 0.4 and 0.8 Å$^{-1}$ for LaFe$_{11.6}$Si$_{1.4}$. The inset highlights their difference that peaks at about 0.5 Å$^{-1}$, which is close to the (100) Bragg peak ($Q \sim 0.54$ Å$^{-1}$) which is crystallographically forbidden.

As the extra magnetic scattering intensity is peaked at 0.5 Å$^{-1}$, detailed constant-$Q$ spectra are investigated at different temperatures for LaFe$_{11.6}$Si$_{1.4}$ (**Fig. 2c**) as well as for LaFe$_{11.2}$Si$_{1.8}$ (**Fig. 2d**). Below $T_C$, the spectra exhibit strong elastic peaks. Just above $T_C$, QENS components are superimposed underneath the elastic peaks. As the temperature continues to rise, the QENS components become broader and weaker. Even so, it is still visible at $1.6T_C$. However, it is apparent that LaFe$_{11.6}$Si$_{1.4}$ exhibits much stronger QENS intensity at the same reduced temperature. Also, the QENS components in LaFe$_{11.6}$Si$_{1.4}$ are more temperature-dependent as the intensity is significantly suppressed when the temperature is changed from 210 to 250 K.

Henceforward, spectral fitting is used to extract accurate information on spin dynamics from the QENS intensity. At 210 K for LaFe$_{11.6}$Si$_{1.4}$, the spectrum at 0.5 Å$^{-1}$ is well reproduced by a combination of a delta function and two Lorentzian functions, which are all convoluted to the instrument resolution function (see **Fig. 3a**). The delta function represents the elastic scattering while the Lorentzian function accounts for the dynamic diffuse scattering containing information on dynamic spin correlations with finite timescales. The wider spectrum is, the faster spin fluctuations exist. This fitting yields a full-width-at-half-maximum ($\Gamma$) of 5.237 and 1.152 meV, which are compared to 5.613 and 1.538 meV at 250 K, respectively. As shown in **Fig. 3b**, a similar fitting is applied to the spectrum of LaFe$_{11.2}$Si$_{1.8}$ at 250 K, which gives rise to $\Gamma$ of 6.309 and 1.099 meV, respectively. Note that these line widths correspond to a typical picosecond timescale. Except for the three spectra mentioned above, others can be well reproduced by a combination of a delta function and one single Lorentzian function. As shown in **Supplementary Fig. 4**, the narrower one disappears with the wider one remained. The temperature- and $Q$-dependent fitting results are summarized in **Supplementary Fig. 5**.

Based on the precise separation of scattering intensity presented above, we sum up the intensity of elastic scattering and the wider QENS component for LaFe$_{11.6}$Si$_{1.4}$ at 210 and 250 K as well as for LaFe$_{11.2}$Si$_{1.8}$ at 250 K, respectively, while all components for spectra at other temperatures are used. The results are plotted as a function of $Q$, which are fitted to equation $S(Q) = M(0)^2 \frac{(1/\xi_{PM})^2}{(1/\xi_{PM})^2 + Q^2}$ to obtain the PM correlation length ($\xi_{PM}$), where $M(0)^2$ is the static susceptibility in units of $\mu_B^2$ (ref. 20). Two examples are shown in **Fig. 3c**. Similar fitting is applied to the data at other temperatures and the obtained temperature dependence of $\xi_{PM}$ is summarized in **Fig. 3d**. As the nearest neighboring Fe-Fe distance is about 2.5 Å in this system, the obtained $\xi_{PM}$ suggests that magnetic correlations are localized around the nearest neighboring Fe atoms at 1.6$T_C$. At a given reduced temperature, it is evident that LaFe$_{11.6}$Si$_{1.4}$ has a slightly smaller $\xi_{PM}$ than LaFe$_{11.2}$Si$_{1.8}$.

Unlike the PM scattering diverging towards lower $Q$, the intensity of the narrower Lorentzian function shows a well-defined peak at $Q \sim 0.5$ Å$^{-1}$, as shown in **Fig. 3e and f**. This position is close to the crystallographically forbidden Bragg peak (100) and the $Q \sim 0.5$ Å$^{-1}$ peak represents ultrafast FM fluctuations[21]. This peak is fitted to a Gaussian function, yielding FM correlation length ($\xi_{FM}$) of 12.2 and 21.0 Å, respectively. It is noted that the FM fluctuations are more remarkable in LaFe$_{11.6}$Si$_{1.4}$ as compared in **Fig. 3e** and **f**. Moreover, at 1.35$T_C$ the FM fluctuations are robust in LaFe$_{11.6}$Si$_{1.4}$ (at 250 K) whereas those become undetectable in LaFe$_{11.2}$Si$_{1.8}$ (at 282 K) within the present experimental resolution. As a result, the extra intensity shown in **Fig. 2b** turns out to originate from FM fluctuations with the timescale of a few picoseconds. Here, we emphasize the uniqueness of the FM fluctuations observed here, compared to the common critical magnetic scattering. Just above magnetic ordering temperatures, magnetic diffuse scattering is often observed to be centered at positions of magnetic Bragg peaks and appears as a QENS signal imposed underneath the elastic scattering[22]. The spectral weight of diffuse scattering is transferred to magnetic Bragg peaks and spin-wave excitations when the systems are magnetically ordered. In contrast, the FM fluctuations are only found in the position of Bragg peak (100). It can be seen in **Supplementary Fig. 6** that the spectra at (200), the first Bragg peak in this system, are dominated by ordinary elastic peaks.

The comparative QENS study on LaFe$_{11.6}$Si$_{1.4}$ and LaFe$_{11.2}$Si$_{1.8}$ suggests that the FOMT is most likely dominated by the spin degree of freedom through picosecond FM fluctuations preceding the transition. To justify such a dominating role, we directly probe the phase transition dynamics in the time domain taking advantage of ultrafast XPD. We focus on the optically induced phase transition and subsequent recovery from sub-ps to 100 ps after the photoexcitation. If the FM fluctuations are indeed the driving force, the FOMT would take place on a similar timescale.

Our ultrafast XPD study was carried out at the SPring-8 Angstrom Compact free-electron LAser (SACLA). As the ultrafast XPD setup is constrained to room temperature, we select the Co-doped sample LaFe$_{10.6}$Co$_{1.0}$Si$_{1.4}$, which undergoes a FM-PM FOMT at about 304 K accompanied by a dramatic negative thermal expansion ($\Delta V/V \sim 1.5\%$, where $V$ is the unit cell volume) with the cubic symmetry remained[13]. An 800 nm laser is used to pump the material across the phase transition. The measurements were completed using the Diverse Application Platform for Hard X-ray diffraction in SACLA (DAPHINS) system[23], where the powder sample was extruded from a nozzle at a fixed rate and the XPD images were acquired in a transmission Debye-Scherrer geometry. The schematic diagram of the experimental setup is shown in **Fig. 4a** and more details are given in **Methods**.

We used an incident laser fluence from 1.26 to 8.74 J cm$^{-2}$ to excite the FOMT. Shown in **Fig. 4b** is the Bragg peak (224) as the strongest peak in the diffraction pattern at selected time delays under the highest laser fluence 8.74 J cm$^{-2}$. The other five Bragg peaks, (024), (135), (026), (246), and (028) are shown in **Supplementary Fig. 7**. Immediately after the laser excitation, the (224) Bragg peak is found to shift to higher angles. The peak-shift maximizes at the time delay of 0.9 ps, with the peak subsequently shifting back to lower angles. At 20 ps, the peak position is almost identical to the Bragg peak (224) position before the optical excitation. Based on the aforementioned six Bragg peaks whose shifts are summarized in **Supplementary Fig. 8**, we derive lattice constants of 11.517 and 11.444 Å before and at 0.9 ps after the laser excitation as shown in **Fig. 4c**. These lattice constants are in agreement with the lattice constants of the FM and PM phases under thermal equilibrium[13]. Thus at 8.74 J cm$^{-2}$, the pump laser induces a full transient phase transition from FM to PM states.

In **Fig. 4b**, the peak position moves continuously between the equilibrium positions that define the FM phase ($Q = 2.6720$ Å$^{-1}$) and PM phase ($Q = 2.6908$ Å$^{-1}$) labeled with the vertical dashed lines, both during the transient phase transition and the recovery that followed. Considering the first-order nature of the phase transition under thermal equilibrium, the lattice constants shown in **Fig. 4b** and **c** reflect the averaged information of a phase-coexistence state. In fact, there is a pronounced broadening of all six Bragg peaks at the earlier stages, indicative of phase-coexistence (**Supplementary Fig. 9**). As such, a more sophisticated fitting has to be employed. To quantitatively describe the phase fraction, the (224) Bragg peaks are fitted to a sum of two Voigt functions centered at $Q = 2.6720$ Å$^{-1}$ for the FM phase and $Q = 2.6908$ Å$^{-1}$ for the PM phase. The fitted peak widths are summarized in **Supplementary Fig. 10.** The FM phase fraction is evaluated by the ratio of the fitted area of the FM peak to the total area of the (224) peak. **Fig. 5a** shows the peak fitting at varying time delay and pump laser fluence. At 1.9 ps, it is evident that higher fluences of the pump laser induce a larger fraction of the PM phase. The fraction of the residual FM phase at the various laser fluences is plotted as a function of laser fluences in **Fig. 5b**. As discussed before, a complete transition occurs at 8.74 J cm$^{-2}$, while under lower fluences, only a fraction of the LaFe$_{10.6}$Co$_{1.0}$Si$_{1.4}$ transforms from the FM phase to the PM phase.

We further determine the timescales relevant to the transient phase transition. The full delay dependence of the FM phase fraction at 8.74 J cm$^{-2}$ is plotted in **Fig. 5c**. The region with smaller time delays is highlighted in **Fig. 5d**. We use the following equation to describe the FM phase fraction as a function of time delay[24,25]

$$f(t) = e^{-t/\tau_d} + (1 - e^{-t/\tau_r}) \quad (1)$$

Here $f(t)$ is the FM phase fraction at time delay $t$. $\tau_d$ and the $\tau_r$ are the characteristic timescales for the suppression and recovery of the FM phase. Our fitting gives an ultrafast $\tau_d$ of 0.3 ps. This timescale is appreciably faster than that reported in similar ultrafast XPD experiments carried out in iron and gold elemental metals[26,27], but well consistent with that of other magnetic systems[28,29]. The ultrafast suppression thus suggests that the phase transition is dominated by factors other than the lattice.

Specifically, the typical size (*l*) of the particles used in our experiment is on the order of ~1 μm (**Supplementary Fig. 11**), which is significantly longer than the optical penetration depth of the 800 nm pump laser. As we are directly observing the evolution of crystalline Bragg peaks from the entire micron-sized particle, for complete suppression of the FM phase at 8.74 J cm$^{-2}$, the hot electrons excited by the pump laser have to travel from the optically excited region throughout the sample at the Fermi velocity to initiate the phase transition. We estimate a Fermi velocity to be ~ $10^6$ m s$^{-1}$ based on $l/\tau_d$, consistent with the typical Fermi velocities in metals[30]. The recovery timescale $\tau_r$ is 4.3 ps, giving rise to a velocity of ~ $10^5$ m s$^{-1}$, which is much larger than sound speeds of a solid so that it should correspond to the timescale of electron-phonon coupling, where the hot electrons and spins dissipate energy into the lattice. To sum, the ultrafast timescale of 0.3 ps amounts to a direct evidence that the electron degree of freedom, rather than the lattice, is responsible for the phase transition.

The phase transition dynamics are established in both frequency and time domains and the FOMT is revealed to be electronic through FM fluctuations. It has been known for several decades that a first-order phase transition occurs when fluctuations are strong enough[31,32]. Recently, fluctuation-induced first-order phase transitions are widely observed in several different systems, like helimagnet MnSi[33] as well as frustrated antiferromagnets $Gd_2Sn_2O_7$ (ref. 34) and $Mn_5Si_3$ (ref. 35). Hence, magnetic fluctuations might be a universal driving force of FOMTs. More specifically, we point out that the obtained conclusion is probably applicable to similar magnetocaloric systems that exhibit FOMTs, like the $(Mn,Fe)_2AsP$ system[36,37].

In summary, we have studied the $LaFe_{13-x}Si_x$ system combining QENS and ultrafast XPD. The comparative QENS study on $LaFe_{11.6}Si_{1.4}$ and $LaFe_{11.2}Si_{1.8}$ indicates the ultrafast FM fluctuations with a typical timescale on the order of 1 ps are relevant to the FOMT, while both compositions display significant phonon softening at the transitions. The dominating part of the electronic degree of freedom is further confirmed on LaFe$_{10.6}$Co$_{1.0}$Si$_{1.4}$ in its ultrafast phase transition process as found in classical magnetic materials. The findings in this model system demonstrate that a universal scenario might be established based on robust FM fluctuations for FOMTs, which are the central issue of giant magnetocaloric materials.

**Acknowledgments**

The work conducted in the Institute of Metal Research was supported by the Ministry of Science and Technology of China (Grant no. 2020YFA0406001), the Key Research Program of Frontier Sciences of Chinese Academy of Sciences (Grant no. ZDBS-LY-JSC002), the Liaoning Revitalization Talents Program (Grant no. XLYC1807122) and National Natural Science Foundation of China (Grant no. 51771197). The work conducted in the Institute of Physics was supported by the Ministry of Science and Technology of China (Grant no. 2019YFA0704900) and the National Natural Science Foundation of China (Grant no. U1832219). Yanna Chen and M. B. were supported by the Institute for Catalysis in Energy Processes (ICEP) under U.S. DOE (Grant no. DE-FG02-03ER15457). The analysis and interpretation of the experiment at SACLA were supported in part by the U.S. Department of Energy, Office of Science, Basic Energy Sciences, Materials Science and Engineering Division, under Contract No. DE-AC02-06CH11357. We acknowledge beam time awarded by ANSTO (Proposal no. P7867) and SACLA (Proposal nos. 2019A8092 and 2019B8077). The authors thank J. S Gardner and E. Brück for the discussion.

**Author Contributions**

B.L. proposed the project. Zhao Zhang was co-supervised by B.L. and Zhidong Zhang. H.Z., F.X.H., and B.G.S. synthesized the samples and characterized the phase purity as well as magnetizations. X.Y. and C.C. performed TEM measurements. Zhao Zhang conducted SEM measurements. Yanna Chen analyzed the TEM and SEM data. Zhao Zhang, D.Y., R.M., C.Y. and B.L. conducted the QENS measurements. Zhao Zhang and B.L. analyzed the QENS data. Yanna Chen, Zhao Zhang, Zhe Zhang, S.O., K.T., M.S., O.S. and M.B. carried out the XFEL experiments. Yanna Chen and Yue Cao reduced and analyzed the XFEL data. B.L. wrote the manuscript with inputs from all authors.

**Data availability**

The data of the present study are available from the corresponding author upon reasonable request.

**Competing interests**

The authors declare no competing interests.

**Additional information**

Supplementary information is available for this paper at XXX.

**Methods**

**Samples and magnetizations**

$LaFe_{11.6}Si_{1.4}$, $LaFe_{11.2}Si_{1.8}$, and $LaFe_{10.6}Co_{1.0}Si_{1.4}$ were prepared using an arc-melting method followed by a long time annealing to reduce the impurity phase of $\alpha$-Fe, as previously described[6]. The obtained materials for the present study were of single-phase without Fe impurity identified in lab X-ray diffraction measurements. Scanning electron microscopy (SEM) (SUPRA 55-VP FEGSEM, Zeiss) and transmission electron microscopy (TEM) (JEM-2100F, JEOL) observations were conducted on powder samples of $LaFe_{10.6}Co_{1.0}Si_{1.4}$ for determining the grain sizes. Magnetization measurements were performed on a SQUID magnetometry (Quantum Design MPMS XL) to determine the transition temperatures, as shown in **Supplementary Fig. 1**. In terms of the temperature dependencies of magnetization at 100 Oe, $T_C$ for $LaFe_{11.6}Si_{1.4}$, $LaFe_{11.2}Si_{1.8}$, and $LaFe_{10.6}Co_{1.0}Si_{1.4}$ are determined to be 184, 208, and 304 K, respectively.

**Neutron scattering**

The neutron scattering experiments were conducted on the time-of-flight neutron spectrometer Pelican at ANSTO[38]. The instrument was configured for incident neutron energy of 3.7 meV, with an energy resolution of 0.135 meV at the elastic line. The powder samples of $LaFe_{11.6}Si_{1.4}$ and $LaFe_{11.2}Si_{1.8}$ were loaded into an annular aluminium can with a sample thickness of 1 mm. The experiments were performed at 100, 160, 210, 250, 275 and 300 K for $LaFe_{11.6}Si_{1.4}$ and 100, 180, 250, 282, 300 and 340 K for $LaFe_{11.2}Si_{1.8}$. A background spectrum from an empty can was collected under the same conditions as the sample measurements. The instrument resolution function was measured on a standard vanadium can at 300 K. The spectrum of the vanadium standard was also used for detector normalization. All data reduction and manipulation, including background subtraction and detector normalization, were done using the Large Array Manipulation Program (LAMP)[39]. The constant-$Q$ data were fitted with the PAN module in Date Analysis and Visualization Environment (DAVE)[40].

**Ultrafast XPD with XFEL**

The XFEL experiments were carried out at BL03, SACLA in Japan[41]. The incident energy of XFEL beam was 15 keV (the wavelength is 0.827 Å) with a pulse repetition rate of 30 Hz. The beam size is 100 μm in both horizontal and vertical directions. A pump laser was generated from a Ti:sapphire and an amplifier was added to boost the pulse energy[42]. The pump laser has a wavelength of 800 nm and a frequency of 30 Hz. The pulse duration is ~30 fs. The pump laser energy was tuned from 143 to 992 μJ. At samples, the laser fluences were evaluated to be 1.26 to 8.74 J cm$^{-2}$ based on the circular beam area (diameter: 100 μm) and an effective factor (69.2%) imposed by the beam intensity distribution. The experiments were carried out on the DAPHINS system with a grease-jet injector and multi-port charge-coupled devices (MPCCDs)[23]. For the measurements, LaFe$_{10.6}$Co$_{1.0}$Si$_{1.4}$ powders with particle sizes smaller than 50 μm were mixed uniformly with a highly viscous DATPE grease[43]. The DATPE grease is made of 25 wt.% dextrin palmitate (Rheopearl KL2, Chiba Flour Milling Co.) dispersed in dialkyl tetraphenyl ether oil (DATPE, S-3230, MORESCO). The mixture was filled into the DAPHNIS injector (inner diameter: 100 μm) and flowed down at a flow rate of 0.024 ml min$^{-1}$ to provide fresh samples. The injected sample flow was perpendicular to the XFEL beam and the pump laser. Both XFEL beam and the pump laser beam were focused on the same region of the flowing sample. The diffraction images were recorded shot-by-shot using MPCCDs and ~1000 shots were accumulated for one data point to obtain clear ring patterns. The main contribution to the effective time resolution is the arrival timing jitter between XFEL and optical laser pulses, which is 256 fs[44]. It was taken as a resolution of the decay time and all measurements were accordingly corrected.

All the diffraction images were firstly background-subtracted using dark-field corrections. The diffraction image of the standard Si sample (NIST standard reference material 640c) was used for the calibration of the distance between sample and MPCCDs (51.8 mm), beam center on MPCCDs, and detector orientation angles. Fixing parameters above using PyFAI package in Python program[45], azimuthal integration of the measured diffraction image was converted to one-dimensional intensity data as a function of scattering vector ($Q$). Six independent Bragg peaks of the $LaFe_{10.6}Co_{1.0}Si_{1.4}$ were used to calculate the average value of lattice constant $a$. By doing so, the lattice constants $a$ for the FM phase and PM phase were both confirmed. Their corresponding $Q$ values were fixed in a two-peak fitting of the (224) profiles to Voigt functions after subtracting a linear background. The fitted areas were used to evalueate the phase fractions, which were fitted to function (1) to obtain timescales. All the fitting was performed using the *lmfit* package for the nonlinear least-squares fitting in Python. For the timescale fitting, the starting time of phase transition was unknown at first. A starting time ($t_0$) in the function (1) was included for the fitting of the phase fraction changing with time delay at 8.74 J cm$^{-2}$. The fitted $t_0$ is equal to 0.12 ps. Finally, all the time delays shown here were corrected with $t_0$. The timescales were fitted after this time correction.

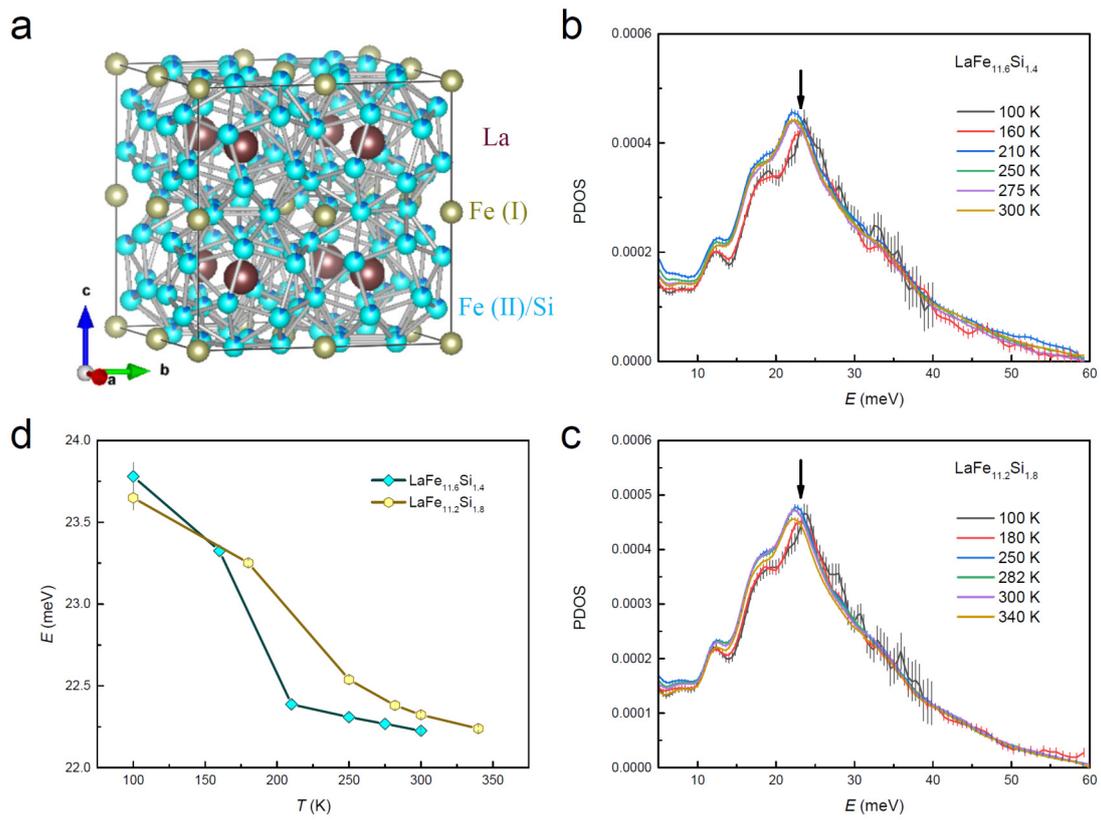

Fig. 1

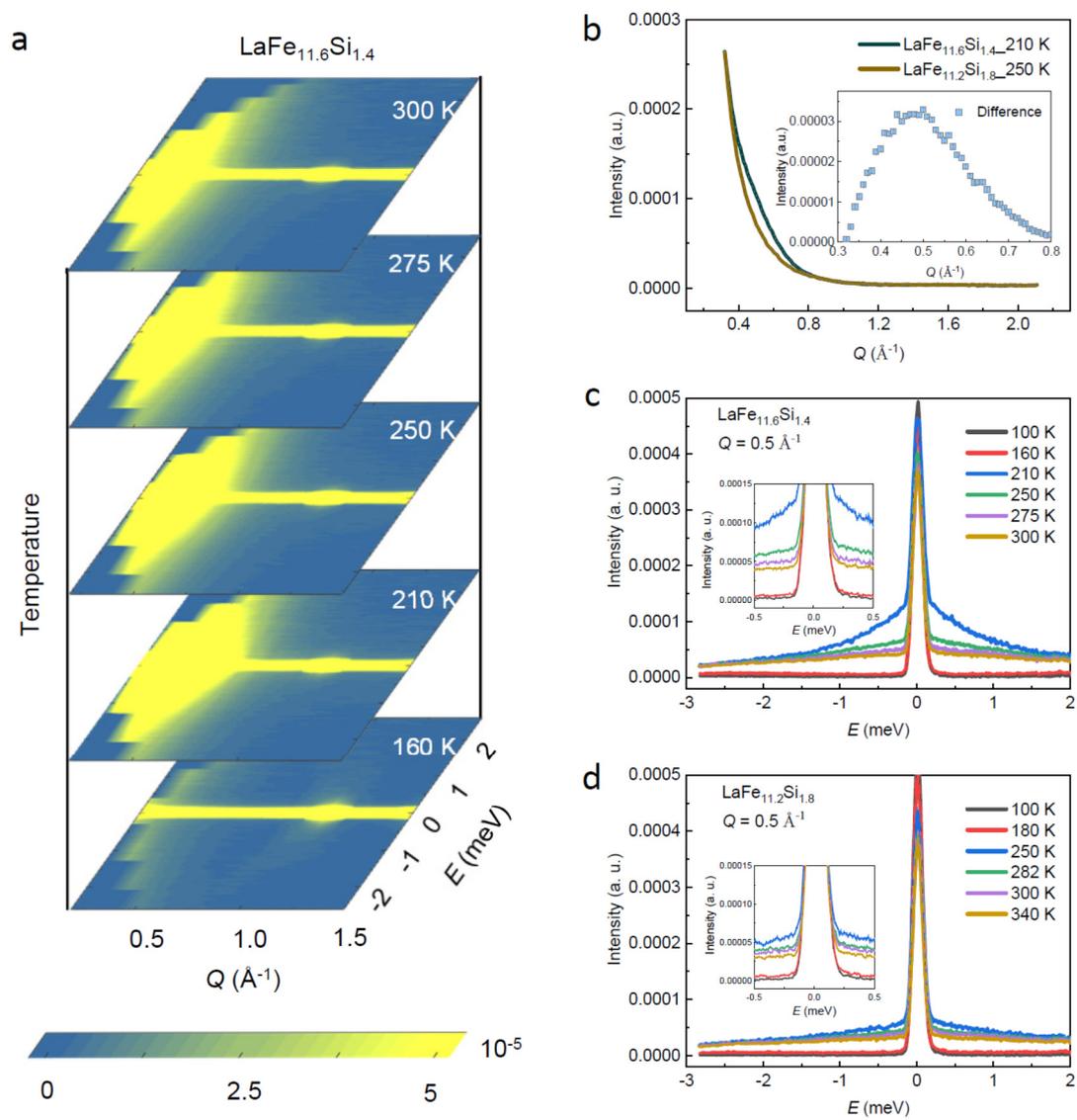

Fig. 2

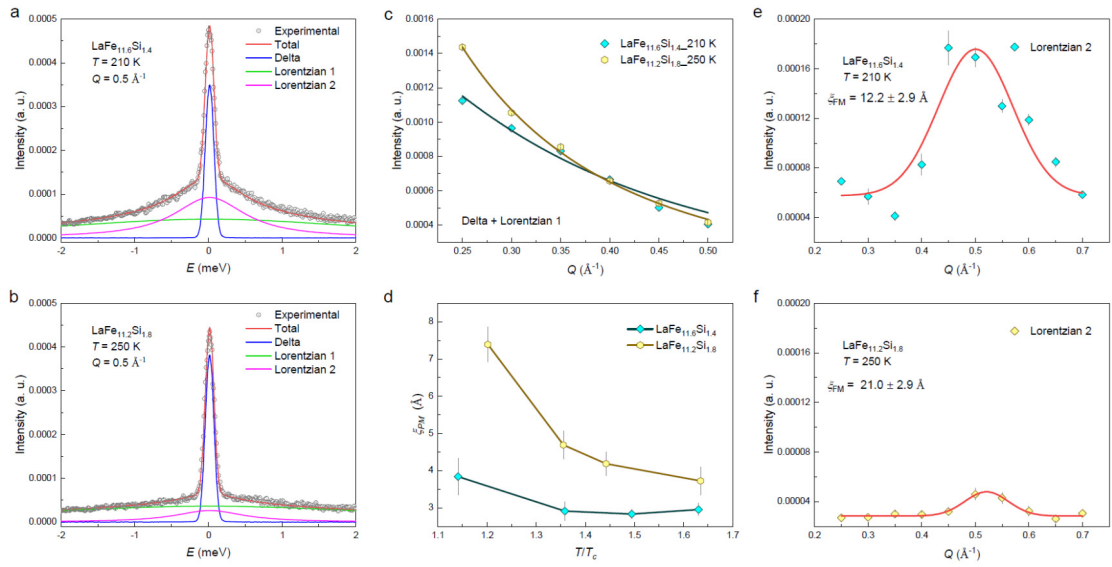

Fig. 3

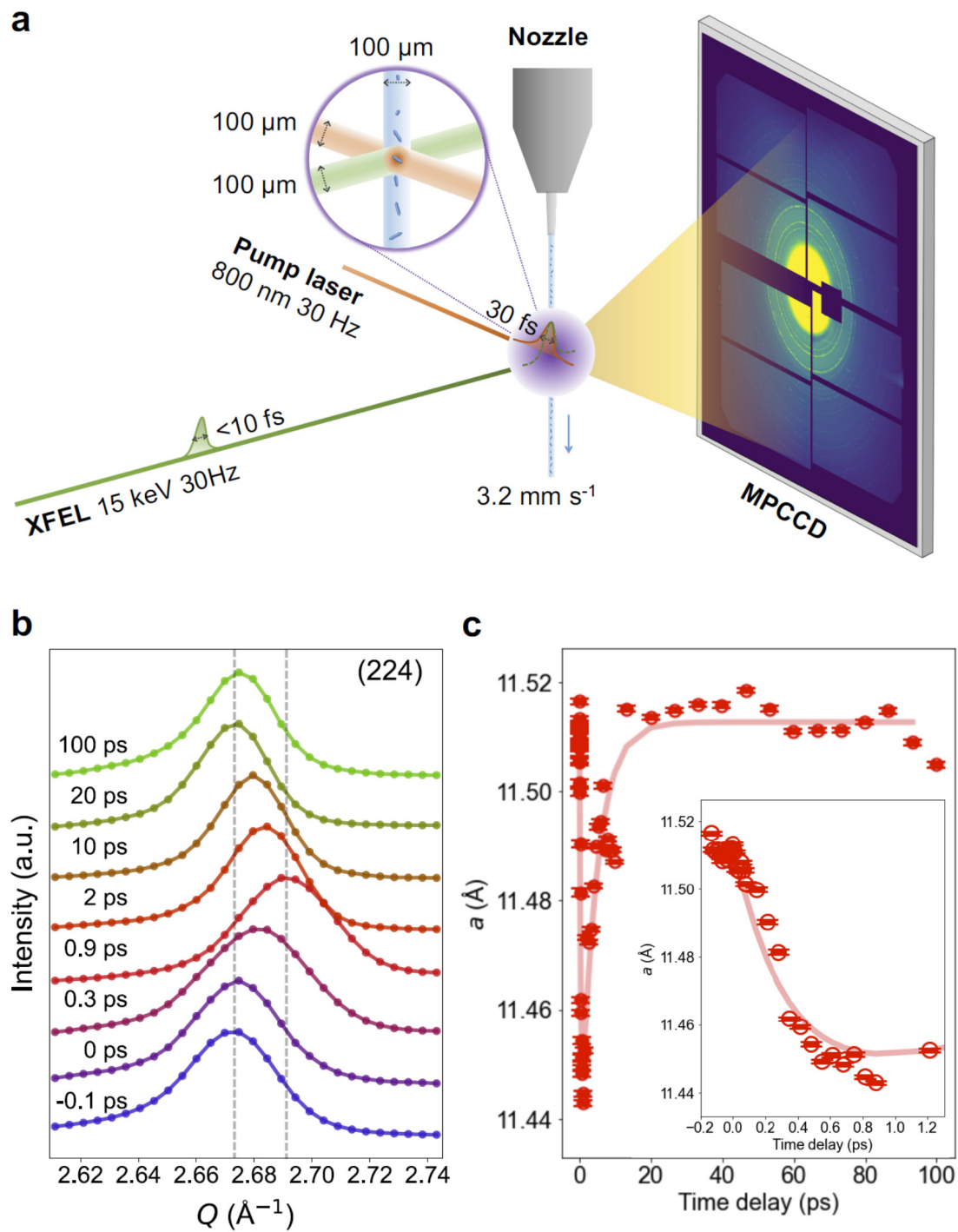

Fig. 4

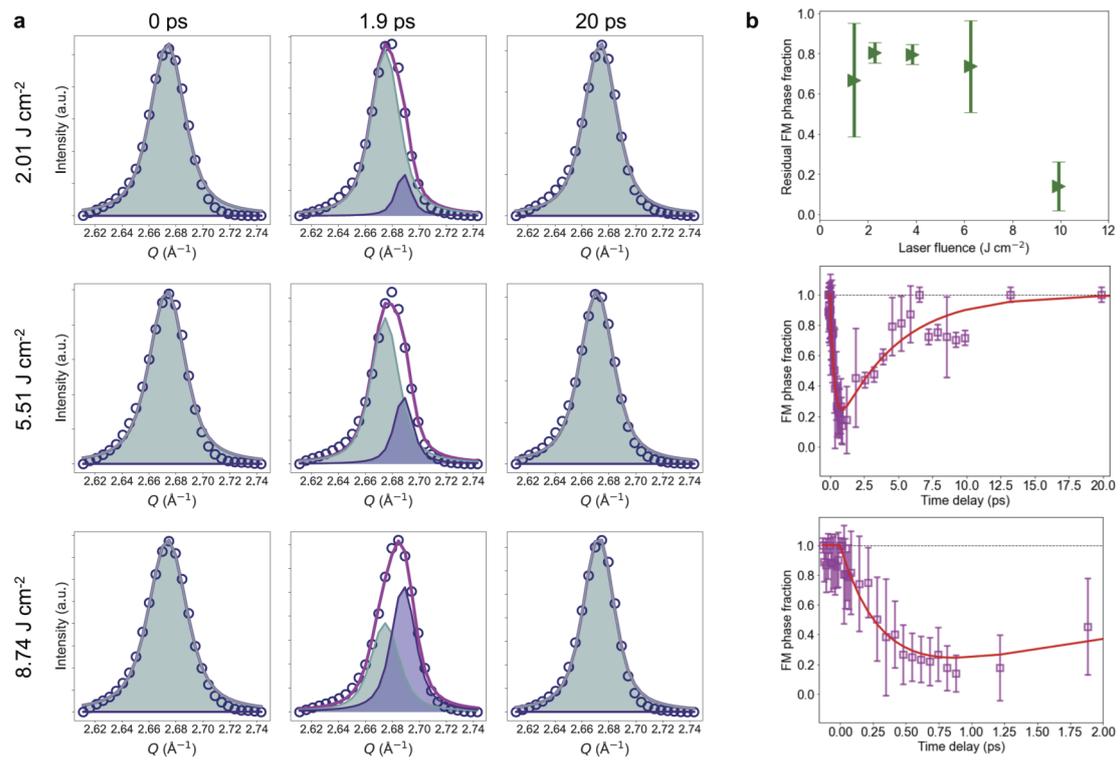

Fig. 5

# Figure captions

**Figure 1 Crystal structure and PDOS. a.** Crystal structure of the $LaFe_{13-x}Si_x$ system. **b,c.** PDOS of $LaFe_{11.6}Si_{1.4}$ and $LaFe_{11.2}Si_{1.8}$ at selected temperature across the phase transitions. The peak positions at about 25 meV are pointed out by arrows. **d.** Temperature dependencies of the peak positions arrowed in **b** and **c**.

**Figure 2 Magnetic diffuse scattering. a.** Dynamic structure factor $S(Q,E)$ of $LaFe_{11.6}Si_{1.4}$ at different temperatures. For $LaFe_{11.2}Si_{1.8}$, refer to **Supplementary Fig. 3**. **b.** Intensity integrated at $-2.55 \leq E \leq -0.15$ meV and $0.15 \leq E \leq 2.6$ meV for $LaFe_{11.6}Si_{1.4}$ at 210 K and for $LaFe_{11.2}Si_{1.8}$ at 250 K. The inset shows their difference. **c,d.** QENS spectra at $Q = 0.5$ Å$^{-1}$ at different temperatures for $LaFe_{11.6}Si_{1.4}$ and $LaFe_{11.2}Si_{1.8}$. The insets highlight the low-$E$ regions.

**Figure 3 Ultrafast FM fluctuations. a,b.** Spectral fitting at $Q = 0.5$ Å$^{-1}$ for $LaFe_{11.6}Si_{1.4}$ at 210 and $LaFe_{11.2}Si_{1.8}$ at 250 K. Each component is highlighted. **c.** $Q$ dependencies of intensity for $LaFe_{11.6}Si_{1.4}$ at 210 K and $LaFe_{11.2}Si_{1.8}$ at 250 K (Delta plus Lorentzian 1). The lines represent the fitting described in the main text. **d**. PM correlation length ($\xi_{PM}$) determined by fitting as shown in **c**. **e,f**. $Q$ dependencies of the intensity of the Lorentzian 2 components for $LaFe_{11.6}Si_{1.4}$ at 210 K and $LaFe_{11.2}Si_{1.8}$ at 250 K. The lines represent the fitting to Gaussian functions yielding the FM correlation length ($\xi_{FM}$) as labeled.

**Figure 4 Ultrafast XPD on $LaFe_{10.6}Co_{1.0}Si_{1.4}$. a.** Experimental configuration for XFEL. The powder sample mixed with grease was extruded from a nozzle at a fixed velocity. XFEL Beam (15 keV, 30 Hz, 0.827 Å) and pump laser (800 nm, 30 Hz) were both focused on the flowing powders. The diffraction patterns were acquired by MPCCDs. **b.** Bragg peak (224) at a laser fluence of 8.74 J cm$^{-2}$ at selected time delays. The vertical dash lines highlight the peak positions of the FM and PM phases. **c**. Lattice

constant evolution upon excitation at 8.74 J cm$^{-2}$. The red line is a guide to the eye. The inset highlights the region of smaller time delays.

**Figure 5 Ultrafast phase transition process of LaFe$_{10.6}$Co$_{1.0}$Si$_{1.4}$. a.** (224) Bragg peak at different incident laser fluences and time delay. The circles are the data points while the solid lines are the fitted profiles. The FM contribution is marked in light green while the PM one is marked in light blue. **b.** The fraction of residual FM phase at different laser fluences. **c.** The FM phase fraction as a function of time delay at the laser fluence of 8.74 J cm$^{-2}$. **d.** The zoomed region of smaller time delays of **c**.

# Supplementary information

**Supplementary Figures 1 – 11**

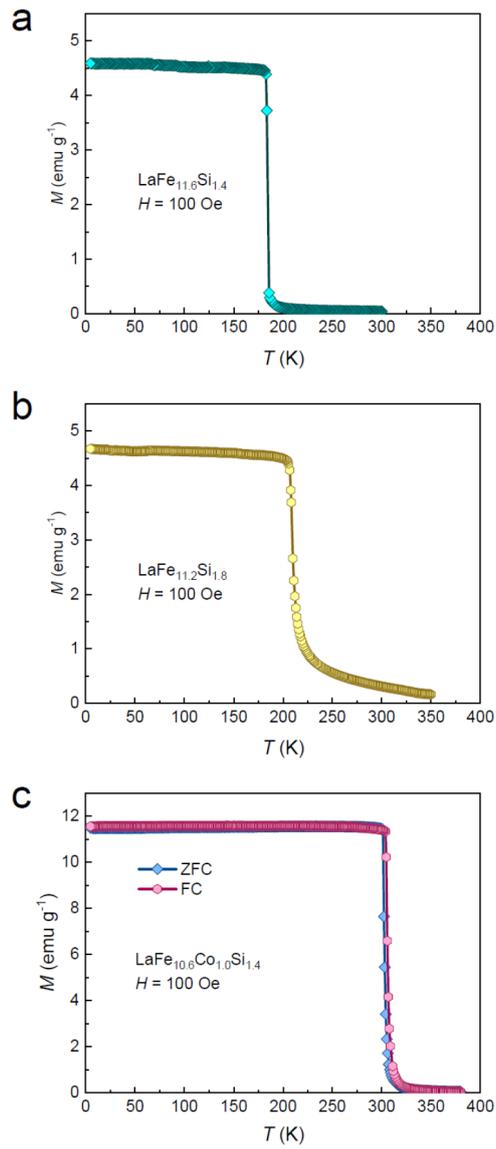

**Supplementary Fig. 1 Temperature dependence of magnetization of three samples used in the present study. a**. LaFe$_{11.6}$Si$_{1.4}$. **b**. LaFe$_{11.2}$Si$_{1.8}$. **c**. LaFe$_{10.6}$Co$_{1.0}$Si$_{1.4}$.

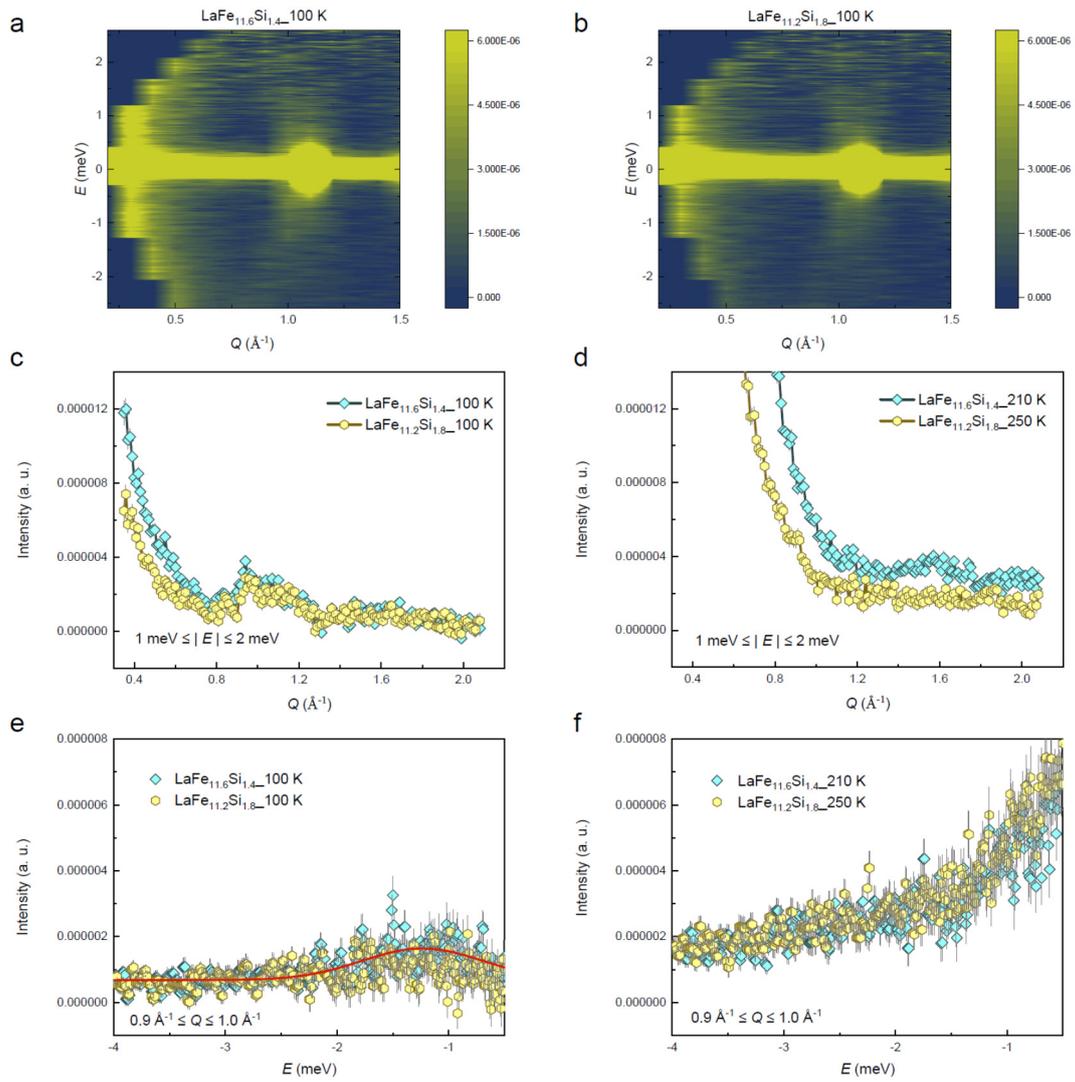

**Supplementary Fig. 2 Spin-wave excitations for LaFe$_{11.6}$Si$_{1.4}$ and LaFe$_{11.2}$Si$_{1.8}$. a**. Contour plot at 100 K for LaFe$_{11.6}$Si$_{1.4}$. **b**. Contour plot at 100 K for LaFe$_{11.2}$Si$_{1.8}$. **c**. Constant-$E$ cutting at 100 K. **d**. Constant-$E$ cutting above $T_C$. **e**. Constant-$Q$ cutting at 100 K. **f**. Constant-$Q$ cutting above $T_C$. It is evident that the excitations disappear above $T_C$, which confirms their magnetic nature.

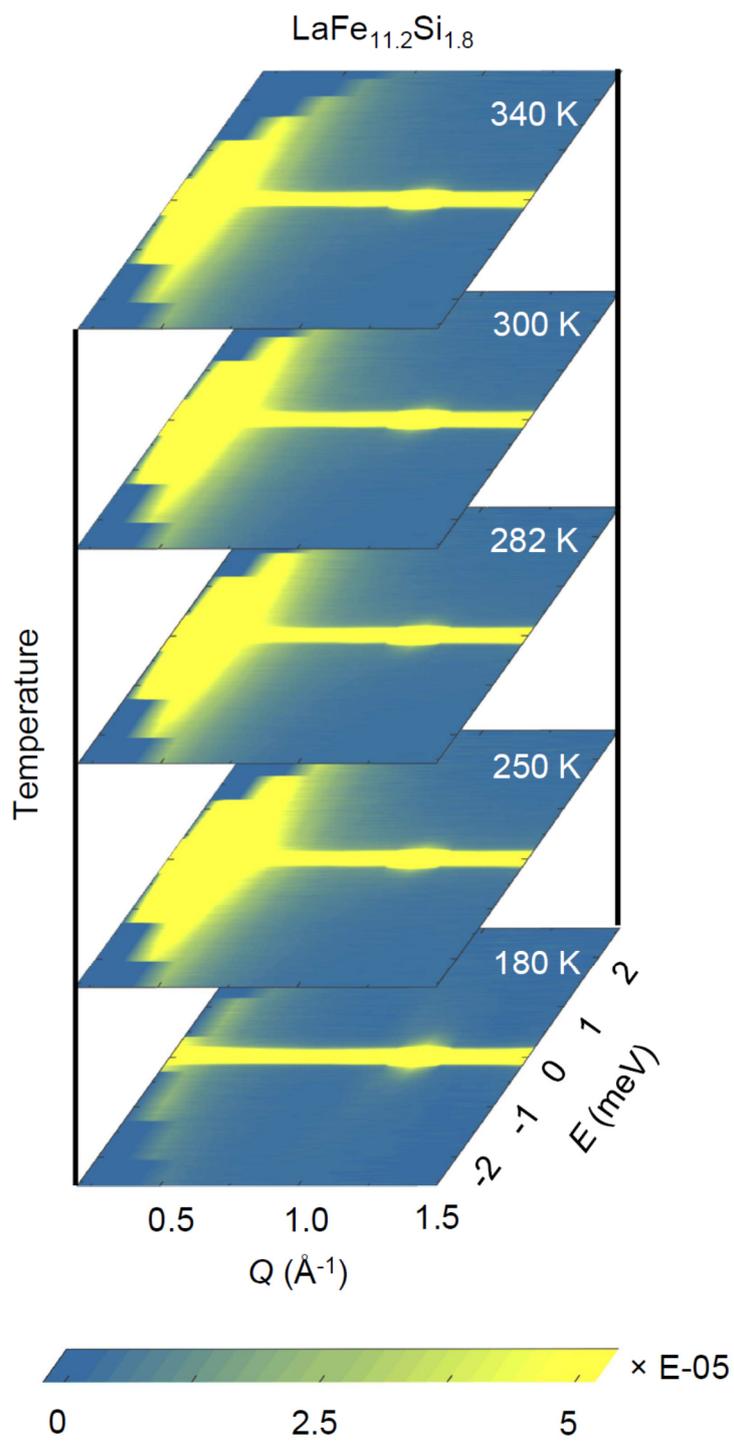

**Supplementary Fig. 3 Dynamic structure factor $S(Q,E)$ obtained in QENS measurements on $LaFe_{11.2}Si_{1.8}$.**

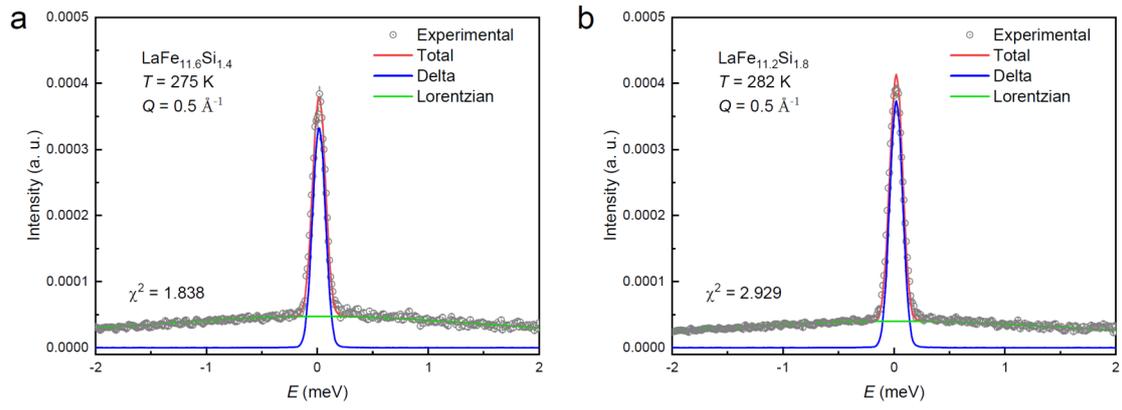

**Supplementary Fig. 4 Fitting of the QENS spectrum at 0.5 Å⁻¹ with one Lorentzian function. a**. LaFe$_{11.6}$Si$_{1.4}$ at 275 K. **b**. LaFe$_{11.2}$Si$_{1.8}$ at 282 K.

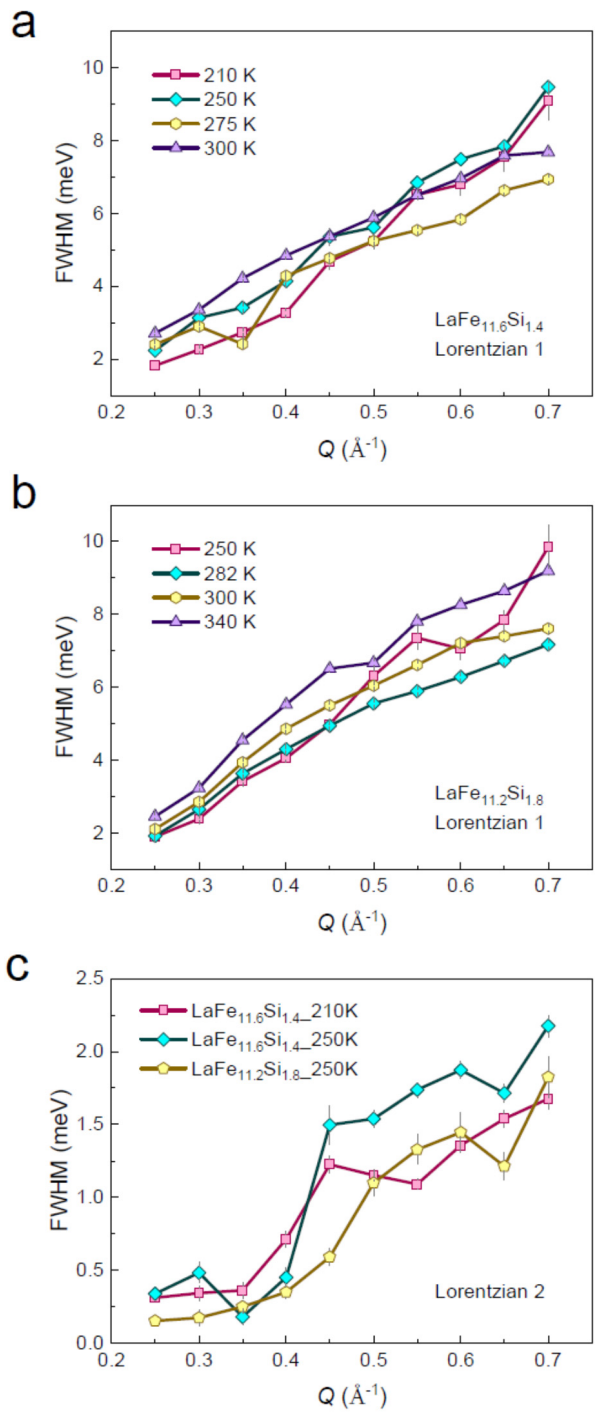

**Supplementary Fig. 5 Full width at half maximum (FWHM) derived from spectral fitting as a function of $Q$. a**. Lorentzian 1 component of $LaFe_{11.6}Si_{1.4}$. **b**. Lorentzian 1 component of $LaFe_{11.2}Si_{1.8}$. **c**. Lorentzian 2 component of $LaFe_{11.6}Si_{1.4}$ and $LaFe_{11.2}Si_{1.8}$.

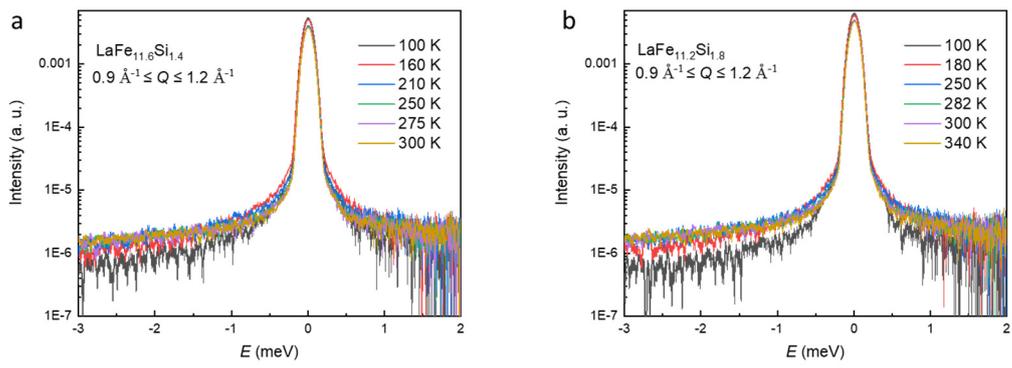

**Supplementary Fig. 6 QENS spectra around the (200) Bragg peak at different temperatures for LaFe$_{11.6}$Si$_{1.4}$ (a) and LaFe$_{11.2}$Si$_{1.8}$ (b).**

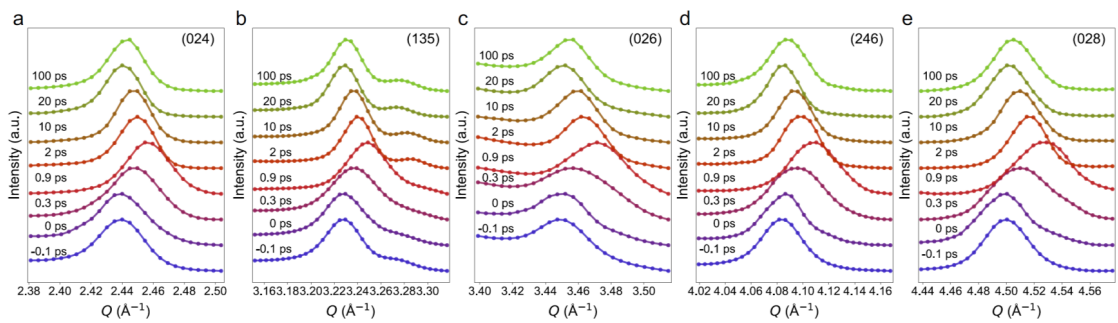

**Supplementary Fig. 7** Azimuthal integrated diffraction profiles for Bragg peaks **a**. (024), **b**. (135), **c**. (026), **d**. (246), and **e**. (028) at the incident laser fluence of 8.74 J cm$^{-2}$.

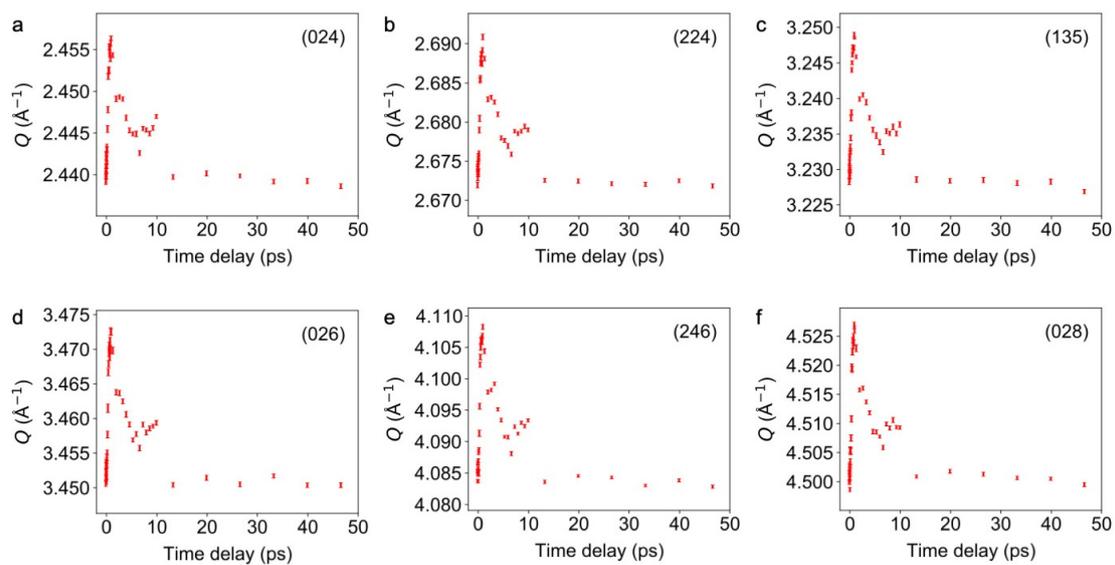

**Supplementary Fig. 8** Peak positions as a function of time delay at the incident laser fluence of 8.74 J cm$^{-2}$ for the six Bragg peaks: **a**. (024), **b**. (224), **c**. (135), **d**. (026), **e**. (246), and **f**. (028).

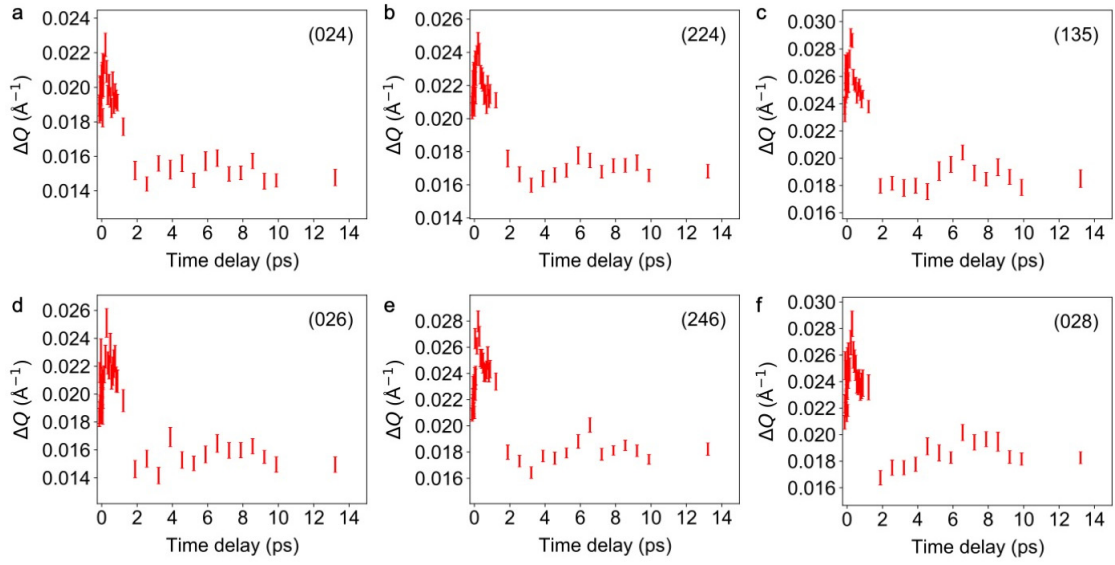

**Supplementary Fig. 9** The full width at half maximum as a function of time delay at the incident laser fluence of 8.74 J cm$^{-2}$ for the following six Bragg peaks: **a**. (024), **b**. (224), **c**. (135), **d**. (026), **e**. (246), and **f**. (028).

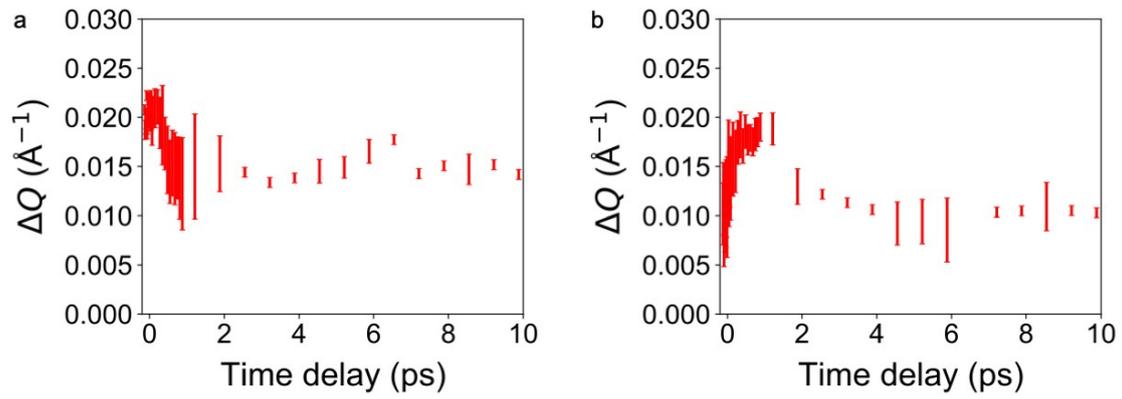

**Supplementary Fig. 10** Fitted full width at half maximum of (224) Bragg peaks for **a**. FM phase ($Q$ = 2.6720 Å$^{-1}$), **b**. PM phase ($Q$ = 2.6908 Å$^{-1}$) at the incident laser fluence of 8.74 J cm$^{-2}$.

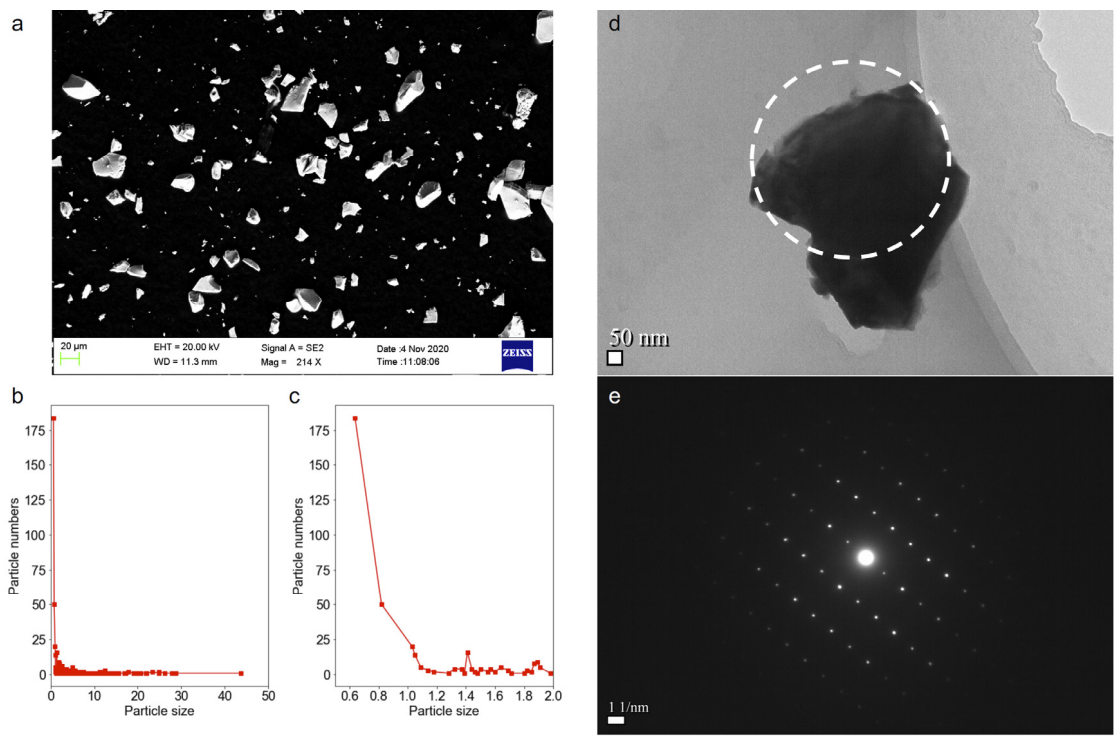

**Supplementary Fig. 11 a**. SEM image of LaFe$_{10.6}$Co$_{1.0}$Si$_{1.4}$ particles. **b**. particle size distribution. **c.** Enlarged region below 2 μm, indicating most of the particles are smaller than 1 μm. **d**. TEM bright-field image of a small particle, where the region for diffraction is highlighted. **e**. Selected area diffraction pattern of this particle, confirming the particle being a single crystal grain.